\documentclass[12pt]{iopart}

\usepackage{graphicx}
\usepackage{bm}

\def\be{\begin{equation}}
\def\ee{\end{equation}}
\def\bea{\begin{eqnarray}}
\def\eea{\end{eqnarray}}

\begin{document}

\title{Cosmological stretching of perturbations on a cosmic string}

\author{Christian Stephan-Otto \dag\, Ken D. Olum \dag\ 
and Xavier Siemens \ddag}
\address{\dag\ Institute of Cosmology,
Department of Physics and Astronomy, Tufts University, Medford, MA
02155, USA}
\address{\ddag\ Center for Gravitation and Cosmology, 
Department of Physics, University of Wisconsin --- Milwaukee, P.O. Box 413,
Wisconsin 53201, USA}

\ead{christian@cosmos.phy.tufts.edu, kdo@cosmos.phy.tufts.edu, 
siemens@gravity.phys.uwm.edu}

\begin{abstract}
We investigate the effects of cosmological expansion on the spectrum
of small-scale structure on a cosmic string.  We simulate the
evolution of a string with two modes that differ in wavelength by one
order of magnitude. Once the short mode is inside the horizon, we find
that its physical amplitude remains unchanged, in spite of the fact
that its comoving wavelength decreases as the longer mode enters the
horizon.  Thus the ratio of amplitude to wavelength for the short mode
becomes larger than it would be in the absence of the long mode.
\end{abstract}

%\preprint{WISC-MILW-03-TH-3}

\maketitle

\section{Introduction} 

Particle physics models that involve symmetry breaking phase
transitions \cite{kibble,vilenkin} as well as some brane-world
scenarios \cite{SarangiTye} predict the formation of topological
defects. Much work has gone into understanding the evolution of
defects (see, for example, \cite{vilenkin}).  Among defects, cosmic
strings have drawn the bulk of the effort because, unlike monopoles
and domain walls, they do not cause cosmological disasters. Indeed,
strings are viable candidates for a variety of interesting
cosmological phenomena such as gamma ray bursts \cite{4},
gravitational wave bursts \cite{5} and ultra high energy cosmic rays
\cite{6}. The detailed predictions for these phenomena, however,
depend sensitively on the spectrum of perturbations present on cosmic
strings.

The result of a phase transition that produces strings is a network of
long strings that stretch across the horizon and a collection of
closed loops. The large scale evolution of a string network is well
understood analytically
\cite{kibble85,bennet86,Martins:2000cs,Martins:1996jp} as well as
numerically \cite{at,bb,as}. Cosmic string networks quickly evolve
toward a ``scaling'' regime in which correlation lengths of long
strings and average sizes of loops scale with the cosmic time $ t
$. This solution is made possible by intersections of long strings
which produce loops. Gravitational radiation causes loops to decay and
so reduces the total length of string in the network. There is much
structure present on the long strings arising from the Brownian
character of the network at formation as well as a build-up of the
sharp edges (kinks) \cite{AllenCald} produced at string intersections.

The lower bound on the size of the structure, also referred to as the
small-scale structure cutoff, is thought to be given by gravitational
back-reaction.  It turns out that a given mode of a perturbation
on a string interacts gravitationally only with a narrow range of
other modes \cite{ks,us}. The range depends on the amplitudes and
wavelengths, and hence the spectrum, of the modes that make up the
perturbations.  The result of this is that the small-scale structure
cutoff depends rather sensitively on the spectrum of perturbations
present on strings.

The spectrum of perturbations, in turn, depends on a combination of
two effects: 1) the cosmological stretching of the primordial
perturbations present on the strings, and 2) the formation and
build-up of kinks due to string intersections.

In \cite{us} a simple model to account for cosmological stretching was
used. In this model all modes with wavelengths smaller than the
horizon are stretched: their wavelengths grow with the scale factor,
while their physical amplitudes are unchanged. In this work we show
that this is only true of modes with wavelengths smaller than, but
comparable to, the horizon.  The evolution of much smaller modes is
more complicated: their physical amplitudes again remain unchanged,
but while they are being stretched by the expansion of the Universe a
change in the arc length of the string acts to decrease their comoving
wavelength.

In Section II we review the motion of strings in expanding
space-time. In Section III we describe the computational scheme as 
well as initial conditions used in our simulations and we discuss the
results. We conclude in Section IV.

\section{Motion of strings in expanding space-time}

If the typical length scale of a cosmic string is large compared to
its thickness, the string can be accurately modeled by a
one-dimensional object.  The equations of motion are obtained from the
Nambu-Goto action, which is proportional to the area swept by the
world-sheet of the string.

We consider the dynamics of strings in an expanding FRW space-time
with a line element $ds^2=a^2(\tau) ( d\tau^2-d{\bf x}^2 )$.  The
comoving spatial coordinates of the string, $\mathbf{x}(\tau,\sigma)$,
are written as functions of the conformal time $\tau$ and a spatial
parameter $\sigma$.

It is convenient to choose a gauge in which the unphysical parallel
components of the velocity vanish
\be \mathbf{\dot x} \cdot \mathbf{x}' = 0.
\ee \label{gauge}
In this gauge the equations of motion of the string are \cite{turok}
\be 
\mathbf{\ddot x} +2\left( \frac{\dot a}{a} \right) \mathbf{\dot x}    
\left( 1 - \mathbf{\dot x}^2\right) = \left( \frac{1}{\epsilon} \right)
\left( \frac{\mathbf{x}'}{\epsilon}\right)'\label{eqmot1}
\ee 
with
\be
\label{epsilon}
\epsilon = \sqrt{ \frac{\mathbf{x'}^2}  {1-\mathbf{\dot x}^2}}
\ee
and
\be
\frac{\dot \epsilon}{\epsilon} 
= -2~\frac{\dot a}{a} ~\mathbf{\dot x}^2 \label{eqmot2}.
\ee
Dots and primes denote derivatives with respect to $\tau$ and $\sigma$
respectively. 

The string's total energy is given by $a(\tau) \mu \int \epsilon d
\sigma$, where $\mu$ is the string's mass per unit length, so that
$\epsilon$ can be thought of as the energy per unit $\sigma$ in
comoving units. There remains some freedom in the choice of
parameterization which we can use to set $\epsilon = 1$ in the initial
conditions.

In flat space-time Equation~(\ref{eqmot1}) reduces to the wave
equation. The general solution can be written as the superposition of
oppositely moving waves
\be
\mathbf{x}(\tau, \sigma)= 
\frac{1}{2} \big[ \mathbf{a}(\sigma-\tau)+\mathbf{b}(\sigma+\tau)\big] 
\label{gensol}
\ee 
with the constraints 
\be
\mathbf{a}'^2= \mathbf{b}'^2= 1
\ee
We can express the functions $\mathbf{a'}$ and $\mathbf{b'}$ in terms 
of $\mathbf{\dot x}$ and $\mathbf{x'}$
\be
\mathbf{a'}(\tau, \sigma)=  
\mathbf{x'}(\tau,\sigma)-\mathbf{\dot x}(\tau,\sigma)
\label{asol}
\ee
and
\be
\label{bsol}
\mathbf{b'}(\tau, \sigma) = 
\mathbf{x'}(\tau,\sigma)+\mathbf{\dot x}(\tau,\sigma).
\ee
Although these equations are derived for the flat space-time case,
they turn out to be useful for the reconstruction of the string
shape in curved space-time.

As in \cite{alexstretch,vilenkin}, we consider perturbations on a
straight static string
\be
\label{sp1}
\mathbf{x}(\tau, \sigma)= \mathbf{c} \sigma + \delta \mathbf{x}(\tau, \sigma)
\ee
with $\mathbf{c}$ constant. In an FRW space-time with power-law
expansion $a(\tau)= \tau^{\alpha}$ the solution to Equations~(\ref{eqmot1})
and (\ref{eqmot2}) for small perturbations is a superposition of waves
of the form
\cite{alexstretch,vilenkin}
\be
\label{sp2}
\delta \mathbf{x}(\tau, \sigma)=\mathbf{A} \tau^{-\nu} 
J_{\nu} (\kappa \tau) e^{i \kappa \sigma}.
\ee
where $\mathbf{A} \cdot \mathbf{c}=0$ and $\nu=\alpha-1/2$.  The
physical wavelength of the perturbations is given by
\be
\label{sp3}
\lambda=a(\tau)\frac{2\pi}{\kappa}
\ee
and therefore $\kappa \tau=2\pi \tau^{\alpha+1} / \lambda \sim t/
\lambda$, is approximately the ratio of the horizon to the size of the
mode. The behavior that follows from Equation~(\ref{sp2}) is rather
simple.

When the wavelength of the mode is large compared to the horizon,
$\kappa \tau \ll 1$, the small argument expansion for the Bessel function
can be used to show that the comoving amplitude is constant in time,
namely,
\be
\label{alarge}
\delta \mathbf{x} \approx \mathbf{A} \left( \frac{\kappa} {2} \right)^{\nu} 
\frac{ e^{i \kappa \sigma}}{\Gamma(\nu+1)}.
\ee
Therefore both the physical wavelength and amplitude are proportional
to the scale factor and the overall size of the string grows while its
shape remains fixed. 

When the mode is well inside the horizon, $\kappa \tau \gg 1$, the large
argument expansion of the Bessel function can be used to show that the
comoving amplitude goes as the inverse of the scale factor
\be
\label{alarge2}
\delta \mathbf{x} \approx \mathbf{A} \tau^{-\alpha} \sqrt{\frac{2}{\pi\kappa}}
\cos (\kappa \tau -\alpha \pi/2) e^{i \kappa \sigma}.
\ee
Therefore the physical amplitude is constant while the wavelength
grows. This results in a straightening of the mode: the amplitude to
wavelength ratio decreases with the scale factor.

If we ignore any interaction between different perturbations, it is
easy to show \cite{us} that we obtain a power-law spectrum for the
amplitude to wavelength ratios.  Shorter perturbations have
been inside the horizon longer, and therefore have been
proportionately more damped than longer ones.

However, if there are simultaneous perturbations of different
wavelengths, and their amplitudes are not too small, we expect
interactions between them.  The effect on the wavelengths of the
perturbations can be understood without simulation.  Let us consider a
string with two perturbations, one with wavelength much less than the
other.  Suppose that the long excitation is in the form
of a standing wave, so that there are periodically times when it does
not contribute to the string motion.  At such time, we can easily
measure the wavelength of the short excitation which is just the
length of one cycle along the underlying string.

The number of cycles of the short excitation in a comoving box is
unchanged by the expansion of the universe, but the arc length of the
underlying string is decreased because the amplitude of the long
excitation is damped.  Thus the comoving wavelength of the short
excitations decreases, since the same number of them fit on a string 
with shorter comoving arc length.  In terms of physical length,
we expect the short wavelength to still increase, but not as much as
it would have, had the long excitation not been present.

On the other hand, the evolution of the amplitude of the short
excitation is not at all obvious.  One might think that when the
horizon is much larger than the short wavelength, the short excitation
physical amplitude remains unchanged, or one might think that the
amplitude to wavelength ratio of the short excitations evolves in the
same way (going inversely with the scale factor) as if the long
excitations were not present.  To determine which of these ideas is
correct, we turn to numerical simulations.

\section{Numerical simulations of strings in expanding space-time}
\label{simulations}

In our simulations the strings evolve according to discretized
versions of Equations~(\ref{eqmot1}) and (\ref{eqmot2}). Because of its
simplicity we have chosen the high-resolution leapfrog scheme
described in \cite{bb}.

In order to avoid the numerical instabilities of the scheme, the
initial conditions must be such that there are no discontinuities (no
kinks) and the velocity of the string is never close to the speed of
light (no cusps). For our purposes it is sufficient to investigate the 
evolution of an initially static cosmic string, with fixed ends and
two modes present. We will use transverse modes to simplify the retrieval 
of information about their amplitudes.

\begin{figure}[htb]
\centering
\includegraphics[width=9 cm]{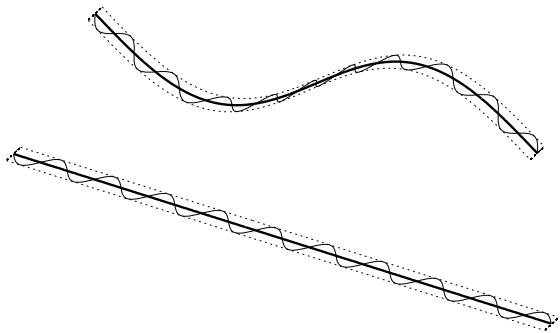}
\caption{
Initial shape of $\mathbf{x}$
according to the ``paper band'' construction. 
The thick solid line depicts the long mode 
$y=A \sin{2\pi z/L}$ in the two-mode case and a straight 
line in the single mode case.}
\label{shape}
\end{figure}

To construct the initial string we start with a single cycle of a
sinusoid \footnote{Note that $y$ is a sinusoid in $z$, not in string
length $\sigma$.  If one uses $\sigma$ instead of $z$ one finds a
maximum amplitude $A = 1$ and with that value parts of the string (in
flat spacetime) move at the speed of light.} $y=A \sin{(2\pi z/L)}$,
let $s$ be its arc length parameter such that $s \in [0,S]$ and let $x
= \alpha \sin{(2 \pi m s / S)}$. The string's shape is defined then by
\be
\mathbf{x}(\tau=0, z)
=(\alpha \sin{\frac{2 \pi m s }{S}}, 
A \sin{\frac{2\pi z}{L}}, z).
\label{init}
\ee
This can be conceptualized as $m$ cycles of a short mode of amplitude
$\alpha$ and wavelength $\lambda=S/m$ that rest on a paper band, which
is then deformed to match its long edge with a simple sinusoid of
amplitude $A$ and wavelength $L$ (see Figure\ \ref{shape}).
Once the values for the string parameters have been chosen and the string 
has been constructed we reparametrize it with respect to the usual arc 
length $\sigma \in [0,\Sigma]$.

To compare the evolution of the amplitude of the short mode in the
previous case to one in which the long mode is not present we evolve
a second set of initial conditions. In this case the static string is
given by
\be
\mathbf{x}(\tau=0, z)
=(\alpha \sin{\frac{2 \pi m z}{S}},0, z)
\label{nobig}
\ee
where $z \in [0,S]$. These initial conditions are such that the length
of the string along the $z$-direction $S$ is the arc-length of the
long mode above.  In terms of the paper band analogy, in this case
the band has been flattened out to lie entirely on the $x$-$z$ plane 
(bottom of Figure\ \ref{shape}).

We have performed a number of simulations to investigate the parameter
space of the initial conditions for the string. Several runs where
carried out for radiation- as well as matter-dominated universes. These
lead to no substantial differences in the behavior. The figures we
show below, which correspond to a universe dominated by radiation, are
representative of all the studied cases.

As in \cite{bb}, we check the accuracy of our results by
numerically evolving $\epsilon$ through the discretized version of
Equation~(\ref{eqmot2}) and calculating it independently from the numerical
results for $\mathbf{\dot x}$ and $\mathbf{x'}$ and
Equation~(\ref{epsilon}).  In our simulations the evolved $\epsilon$ is in
agreement with the value computed from $\mathbf{x'}$ and $\mathbf{\dot
x}$ to within $0.1\%$ for all cases.  The initial number of spatial
oscillations along the string is conserved, meaning that the equations
of motion do not lead to the generation of new modes.

We set the initial ratios of amplitude to angular wavelength,
$\mathcal{E}_l=2\pi A/L$ and $\mathcal{E}_s = 2\pi \alpha/\lambda$,
for the long and short modes respectively, to be unity.  These
quantities characterize the maximum slope of each mode on the string.
We take $m=10$ cycles of the short mode, and choose the initial
amplitude of the long mode to be $A=100$ in arbitrary units.  The
wavelength of the long mode then becomes $L=200\pi$ and the elliptic
integral for the arc length yields $S \approx 764.04$, which gives
$\lambda \approx 76.40$ and $\alpha \approx 12.16$.

For convenience we chose to simulate the string by a series of points
with the comoving spatial distance between each and the next
exactly 1 at the initial time.  To make that possible we decrease
$\alpha$ and thus $\mathcal{E}_s$ by $0.7\%$, obtaining 928 as the
total number of points.

\begin{figure}[htb]
\centering
\includegraphics[width=8.5 cm]{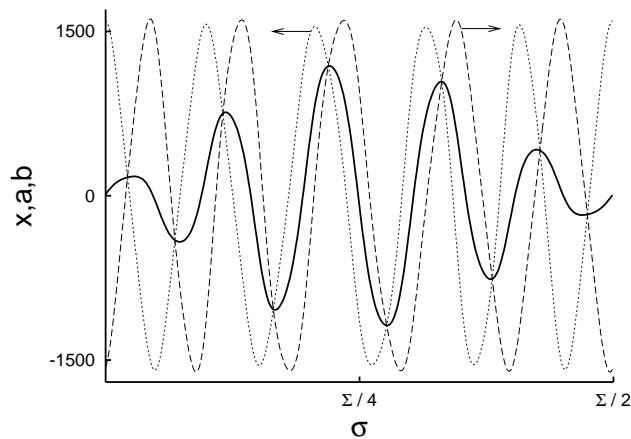}
\caption{ 
Physical $x$-direction perturbations of $\mathbf{x}$ and its traveling 
components $\mathbf{a}$ and $\mathbf{b}$ (solid, dashed and dotted lines 
respectively) as functions of $\sigma$, in a radiation-dominated 
universe. 
Only the first half of the string is shown as the second is its 
repetition.
The arrows tell us that when the snapshot was taken neighboring
peaks in $\mathbf{a}$ and $\mathbf{b}$ were receding from each other, 
indicating a decrease in the physical $\mathbf{x}$ perturbation. 
The snapshot was taken at $\tau=194.6$.
}
\label{abx}
\end{figure}

In order to understand the behavior of the amplitude of the short mode
as a function of time we require a means of extracting its amplitude.
As the simulation progresses, the shape of the short mode can be seen
to vary along the string, as shown in Figure\ \ref{abx}, so that it is
not obvious what its amplitude is.  The reason for this variation is
that the maxima and minima of the oscillations of the short mode are
reached at different times depending on whether they were located at
nodes or anti-nodes of the long mode initially.

To determine the amplitude we compute $\mathbf{a}$ and $\mathbf{b}$ by
numerically integrating Equations~(\ref{asol}) and (\ref{bsol}).  As shown
in Figure\ \ref{abx}, $\mathbf{a}$ and $\mathbf{b}$ approximately
preserve the original comoving shape of the string, except for slight
shifts in the distance separating their nodes, and we can readily
determine the amplitude of the short mode from either of them.  Here
the short mode peaks of $\mathbf{a}$ and $\mathbf{b}$ are closer to
each other for $\sigma \sim \Sigma/4$ than for $\sigma$ either small
or $\sim \Sigma / 2$. Nevertheless, their physical amplitude is
constant along the string.

\begin{figure}[htb]
\centering \includegraphics[width=8.5 cm]{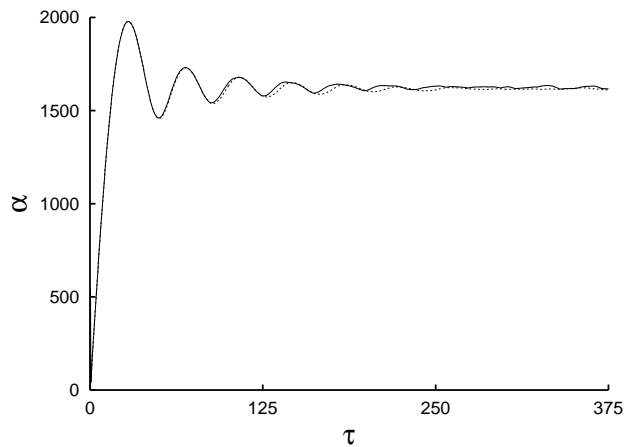} 
\caption{Physical amplitude of the short mode $\alpha$ as a function of
$\tau$ for a string with both long and short modes (solid line) and
for a string with only the short mode (dotted line), in a
radiation-dominated universe.}
\label{asmall}
\end{figure}

In Figure\ \ref{asmall} we show the physical amplitude of the short
mode as a function of $\tau$ for two cases. For one of them the short
mode is the only perturbation on the string, while for the other both
the short and long modes are present.  The starting time of the
simulation is $\tau_i=0.1$ to ensure it is much smaller than both of
the scales defining the string. The corresponding initial value of the
scale factor is $a_i=1$.  The general behavior in the two cases is
the same and is consistent with the solutions presented in Section II as
well the analysis presented in \cite{turok}. Thus the
amplitude of the short mode is insensitive to the presence of a long
one.

The evolution of the physical amplitude of the short mode occurs in
three stages. At first, when the wavelength is longer than the
horizon, conformal stretching prevails and the amplitude grows with
the scale factor, as we expect from Equation~(\ref{alarge}). The second
stage occurs when the short is less than but still comparable to the
horizon size. In this stage the amplitude exhibits
damped oscillatory behavior.  In the final stage the amplitude
becomes constant, as expected from Equation~(\ref{alarge2}).

Although the physical amplitude of the short mode is not affected by
the presence of the long mode, we have argued in Section II that its
wavelength should change as a result of the decreasing arc length of
the string. The evolution of the amplitude to wavelength ratio of
short modes is therefore sensitive to the presence of long modes.

\begin{figure}[htb]
\centering
\includegraphics[width=8.5 cm]{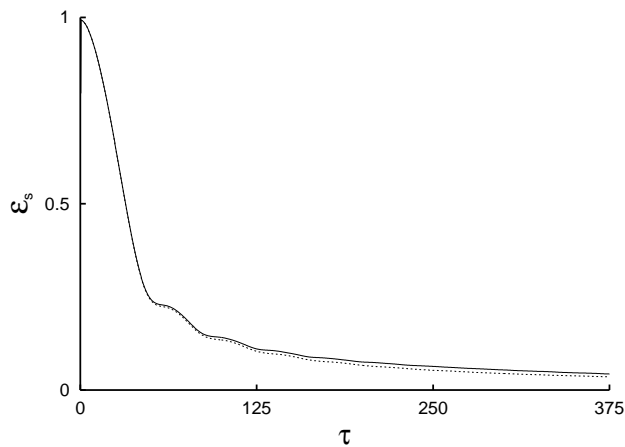}
\caption{Plot of $\mathcal{E}_s$ for a string with long and 
short modes (solid line) and for a string with only the short mode
(dotted line).}
\label{epsilon_s}
\end{figure}

Figure \ref{epsilon_s} shows the evolution of
$\mathcal{E}_s=2\pi\alpha/\lambda$ for the cases shown in Figure\
\ref{asmall}.  The wavelength of the short mode evolves differently
for the two situations presented. For the string with two modes the
comoving arc length of the string decreases with time, causing
$\mathcal{E}_s$ to decrease at a somewhat slower rate than in the the
single-excitation case.

\begin{figure}[t]
\centering
\includegraphics[width=8.5 cm]{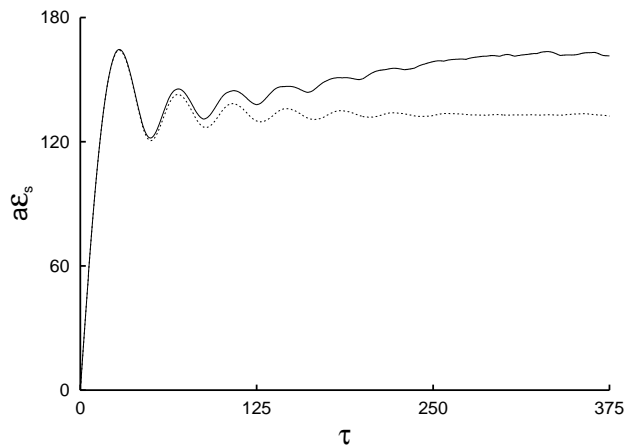}
\caption{Plot of $a(\tau) \mathcal{E}_s$ for a string with long and 
short modes (solid line) and for a string with only the short mode
(dotted line).}
\label{aepsilon}
\end{figure}

To make the discrepancy more evident we show in Figure\ \ref{aepsilon}
the amplitude to wavelength ratio $\mathcal{E}_s$ multiplied by the
scale factor $a$. For the single-mode case the evolution is exactly
the three stage process for the amplitude $\alpha$ described
above, because the wavelength of the short mode $\lambda$
simply scales with the scale factor.  On the other hand, the two-mode
case shows a more complicated evolution process.

At first both modes are larger than the horizon and conformal
stretching prevails; this stage ends with the fall of the short mode into
the horizon at $\tau \sim 10$.  In the second stage the short
perturbation is inside the horizon but the long mode is still
stretching conformally, thus the situation is the same as in the
single mode case.  Once the long mode falls into the horizon its
amplitude is damped, causing the arc length of the string to grow at a
rate smaller than the scale factor. This is observed at $\tau \sim 50$
where $a\mathcal{E}_s$ begins growing, deviating from the single
perturbation case.  The final stage is characterized by the amplitude
of the long mode being negligible compared to its wavelength. Its arc
length is then comparable to its wavelength, which scales with $a$. 
From then on $a \mathcal{E}_s$ remains constant, albeit larger than 
in the single perturbation case.

Although here we have only considered the two mode case, in a
cosmological setting long modes are always entering the horizon. This
means that the fourth stage described above is never reached by a
realistic network of strings and the scaled amplitude to wavelength
ratio (as in Figure\ \ref{aepsilon}) does not stop increasing.

% This works around a bug in Latex where stuck floats cause the white
% space here to be lost.
\mbox{}

\section{Discussion and Conclusions}

Astrophysical and cosmological effects of a cosmic string network
depend sensitively on the small-scale structure present on the
strings.  That small-scale structure exists at the time of string
formation and is further enhanced by intercommutations.  The structure
then evolves by cosmological expansion, and gravitational
back-reaction smooths out all structures smaller than a certain
cutoff.  Gravitational back-reaction at a given scale, however,
depends in turn on the spectrum of excitations at larger scales
\cite{us}.

Thus an accurate understanding of cosmological stretching of string
modes is necessary to understand small-scale structure.  Ref.\
\cite{us} used a simple model which neglected interactions between
different modes and assumed that once modes enter the horizon their
physical amplitudes remain constant while their wavelengths are
stretched.  This model results in amplitude to wavelength ratios that
decrease with decreasing wavelengths. The reason for this is that
short modes have been in the horizon longer and have had more time to
be stretched.

Here we have argued that there is a coupling between modes of very
different wavelengths. While it is true that short modes are being
stretched by the expansion of the Universe, the stretching of long
modes that have recently entered the horizon produces a decrease in
the arc length of the string which acts to decrease the wavelength of
the short modes. The net result is that the wavelengths of short modes
are not stretched as effectively by the expansion of the universe as
the wavelength of long modes.

We have studied the evolution of the amplitudes numerically and found
that, in contrast with the wavelength evolution, the amplitude
evolution does not couple modes of different sizes: Regardless of the
presence of other modes, a mode's physical amplitude remains fixed
once it is well inside the horizon.

Although we have simulated only the case of smooth excitations in
transverse directions, we believe our results are generic.  The effect
on the wavelength of the short mode can be understood analytically
as explained in section II, and depends only on the loss of arc length
in the long mode, and not on its shape or direction.  We found
that the amplitude of the short mode is insensitive to the presence
of the long one, and we expect that principle to be generic.  
When there is a significant difference in wavelength, the long
mode moves the string underlying the short mode as a whole, and we
would not expect the direction of that motion to have any effect on
the short mode evolution.

As compared to the previous simple model we expect the amplitude to
wavelength ratio of the perturbations on cosmic strings to decrease
more slowly as a function of decreasing frequency.  The detailed
consequences for the spectrum and the small-scale structure cutoff will
be examined in detail elsewhere.

\section{Acknowledgments}

We would like to thank Eduardo Calvillo and Alex Vilenkin for helpful
conversations.  C. S.-O. was supported in part by CONACYT.
K. D. O. was supported in part by the National Science Foundation.
X. S. was supported by National Science Foundation grants PHY 0071028
and PHY 0079683.

%\section*{References}

%\begin{thebibliography}{99}

%\end{thebibliography}

\end{document}